\documentclass[aps,prd,twocolumn, reprint, amsmath, amssymb]{revtex4-1}
\usepackage[centering,hmargin=2cm,vmargin=2cm]{geometry}
\usepackage{graphicx}

\newcommand{\td}{\text{d}}
\def\be{\begin{equation}}
\def\ee{\end{equation}}
\def\bea{\begin{eqnarray}}
\def\eea{\end{eqnarray}}

\begin{document}

\title{Black lenses in string theory}

\author{Hari K. Kunduri}
\affiliation{Department of Mathematics and Statistics, Memorial University of Newfoundland, St John's, Canada}

\author{James Lucietti}
\affiliation{School of Mathematics and Maxwell Institute of Mathematical Sciences, University of Edinburgh, King's Buildings, Edinburgh, UK}

\begin{abstract}
We present a new supersymmetric, asymptotically flat,  black hole solution to five-dimensional $U(1)^3$-supergravity which is regular on and outside an event horizon of lens space topology $L(2,1)$. The solution has seven independent parameters and uplifts to a 
 family of 1/8-supersymmetric D1-D5-P black brane solutions to Type IIB supergravity.  
The decoupling limit is asymptotically AdS$_3 \times S^3 \times T^4$, with a near-horizon geometry that is a twisted product of the near-horizon geometry of the extremal BTZ black hole and $L(2,1)\times T^4$, although it  is not (locally) a product space in the bulk.  We show that the decoupling limit of a special case of the black lens is related to that of a black ring by spectral flow, thereby supplying an account of its entropy. Analogous solutions of $U(1)^N$-supergravity are also presented.
\end{abstract}

\maketitle

\section{Introduction}
A significant achievement of string theory has been to provide a microscopic accounting of the Bekenstein-Hawking entropy of supersymmetric black holes~\cite{Strominger:1996sh}.  The black holes are five-dimensional versions of the extremal Reissner-Nordstr\"om solution and include rotating generalisations~\cite{Breckenridge:1996is}.  The black holes have an equivalent description in string theory as configurations of D-branes and their degeneracy for given macroscopic charges can be computed by exploiting supersymmetry.  The decoupling limit of the corresponding black brane solutions possesses a (locally) AdS$_3$ factor. This allows one to appeal to the AdS-CFT duality to provide an alternative explanation for the entropy from the degeneracy of near-horizon microstates in the dual CFT~\cite{Strominger:1997eq}.
\par  
The discovery of black rings revealed that the asymptotic charges are not sufficient to specify a black hole~\cite{Emparan:2006mm}.  However, the black hole microstate arguments typically count the number of states with given charges. 
This did not pose a threat to the original calculations, since in contrast to the spherical black holes, supersymmetric black rings~\cite{Bena:2004de, Elvang:2004ds, Gauntlett:2004qy} have distinct angular momenta.  In fact a microscopic accounting of the entropy of the black ring has been provided by appealing to M-theory~\cite{Cyrier:2004hj}, although a fully satisfactory D-brane argument is lacking~\cite{Emparan:2006mm} (see~\cite{Bena:2004tk} for partial results).

 An important question is whether other families of black holes exist in this context. Recent work has revealed that the classification of asymptotically flat five-dimensional supersymmetric black holes is far from complete~\cite{Kunduri:2014iga, Kunduri:2014kja}.  Furthermore, recent work in the corresponding D-brane CFT has also revealed a rich phase structure~\cite{Bena:2011zw, Haghighat:2015ega}.
 In particular, we constructed the first example of a regular asymptotically flat black hole with lens space topology $L(2,1) = S^3/ \mathbb{Z}_2$~\cite{Kunduri:2014kja}.  The purpose of this note is to generalise and embed these solutions into string theory in order to clarify their microscopic description.   Interestingly, we find that in a special case, the decoupling limit of the corresponding D-brane geometry is related by spectral flow to that of the black ring thus allowing one to appeal to existing microscopic accountings of the entropy~\cite{Cyrier:2004hj, Bena:2004tk}. The general case though remains open.
 
 In section \ref{sec:blens} we present a black lens solution to five dimensional $U(1)^3$-supergravity. In section \ref{sec:D1D5P} we discuss its uplift to a D1-D5-P solution to IIB supergravity and the decoupling limit. In Appendix \ref{appendix} we provide a derivation and a detailed regularity analysis of analogous black lens solutions to $U(1)^N$-supergravity.

 \section{Multi-charge black lenses}
 \label{sec:blens}
 
 \subsection{Supersymmetric solutions}
The bosonic sector of five-dimensional $\mathcal{N}=1, \; U(1)^3$-supergravity is a metric, Maxwell fields $F^i= \td A^i$ and positive scalar fields $X^i$, $i=1,2,3$, obeying $X^1 X^2 X^3=1$. 
A large class of supersymmetric solutions to this theory can be constructed as timelike fibrations over a Gibbons-Hawking (GH) base space \cite{Gauntlett:2004qy}. The GH base is specified by a harmonic function $H$ on $\mathbb{R}^3$ and the  supersymmetric solution is specified by a further 7 harmonic functions $K^i, L_i, M$. In coordinates $(t,  \psi, r, \theta, \phi)$, where $(r,\theta, \phi)$ are spherical polar coordinates on $\mathbb{R}^3$,  the solution is 
\begin{align}
\td s_5^2 &= -f^2 (\td t+ \omega)^2 \nonumber \\ &\phantom{=}+f^{-1} [H^{-1} ( \td \psi+ \chi)^2  +H ( \td r^2 + r^2 \td \Omega_2^2)] \; , \nonumber \\  \nonumber 
A^i &= \frac{1}{3} H_i^{-1} (\td t +\omega) + \frac{1}{2} \left( \frac{K^i}{H} (\td \psi+ \chi) + \xi^i \right) , \\
X^i &= H_i^{-1} (H_1 H_2H_3)^{1/3}  ,\nonumber \\
 f &= \frac{1}{3} \left(H_1 H_2 H_3\right)^{-1/3}\; , \quad  \omega = \omega_\psi ( \td \psi + \chi) + \hat{\omega} \; ,\nonumber  \\
H_i &= L_i+ \frac{1}{24}H^{-1}|\epsilon_{ijk}|K^j K^k , \nonumber  \\
\omega_\psi  &= -\frac{K^1 K^2 K^3}{8H^2} - \frac{3 L_i K^i}{4H} + M   \; ,
 \end{align}
where $\td \Omega_2^2 = \td \theta^2 + \sin^2 \theta \td \phi^2$,  $\epsilon_{ijk}$ is the alternating symbol and $\chi, \xi^i, \hat{\omega}$ are 1-forms on $\mathbb{R}^3$ determined by the harmonic functions up to quadratures \cite{Gauntlett:2004qy}. 
 
 Within this class we have found a family of asymptotically Minkowski, black hole solutions with lens space horizon topology. The construction is straightforward and begins with a multi-centred ansatz of the type studied for soliton geometries~\cite{Bena:2005va, Berglund:2005vb}. 
  The solution is
 \begin{align}
 H &= \frac{2}{r} - \frac{1}{r_1} , \qquad K^i = \frac{k^i}{r} 
  \; ,  \nonumber \\   L_i &= \lambda_i + \frac{\ell_i}{r}  ,\qquad M =   \frac{3 \lambda_i k^i}{4} \left( 1- \frac{a}{r} \right) \; , \nonumber \\
  \chi &= \left[ 2\cos\theta - \frac{r\cos\theta - a}{r_1} \right] \td\phi  \; ,  \nonumber \\
  \hat{\omega} &= - \frac{3 a \lambda_i k^i r \sin^2\theta}{2 r_1 (r_1+r+a)} \td \phi   , \quad \xi^i = -k^i \cos \theta  \td \phi \; ,   \label{blens}
 \end{align}
where $r_1 = \sqrt{r^2+a^2- 2r a \cos \theta }$ is the Euclidean distance from a `centre' in $\mathbb{R}^3$ with Cartesian coordinates $(0,0,a)$  and we assume $a>0$.  

The solution is asymptotically flat $\mathbb{R}^{1,4}$ provided $\lambda_i = 1/3$ and $\Delta \psi = 4\pi$. Indeed, setting $r=\tfrac{1}{4} \rho^2$,  as $\rho \to \infty$
\begin{align}
\td s_5^2 \sim - \td t^2 +  \td \rho^2 + \tfrac{1}{4}\rho^2 \left[ (\td \psi + \cos \theta \td \phi)^2 + \td \Omega_2^2 \right] ,  \label{AF}
\end{align}
 with subleading terms of order $O(\rho^{-2})$.

 The metric and scalars are smooth at $r_1=0$ provided
\begin{align}
\ell_i < - a \lambda_i \; \label{ineq1}
\end{align} 
and  $\Delta \psi= 4\pi$.
Then, the spacetime as $r_1\to 0$ smoothly approaches $\mathbb{R}^{1,4}$. As explained in the Appendix, polar coordinates $(X,\Phi)$ and $(Y, \Psi)$ on the orthogonal 2-planes in $\mathbb{R}^4$ are given by $4 r_1= X^2+Y^2$, $\Phi=\tfrac{1}{2} (\psi+ \phi)$ and $\Psi = \tfrac{1}{2} ( \psi+3 \phi)$. The gauge fields
\begin{align}
A^i &= \frac{ \td t}{3 H_i}+  [ \tfrac{1}{2} k^i+  O(X^2)] \td \Phi - [ \tfrac{1}{2} k^i+ O(Y^2)] \td \Psi \label{gaugecentre}
\end{align}
are thus smooth at $r_1=0$ up to a gauge transformation.

The spacetime has a regular horizon at $r=0$ provided
\bea
 &&h_i \equiv \ell_i+ \frac{1}{48} | \epsilon_{ijk} | k^j k^k >0 ,  \label{h0ineq} \\  
&&\beta \equiv \frac{3}{4} k^i \left(  \ell_i+ 2a \lambda_i+ \frac{1}{72} | \epsilon_{ijk} | k^j k^k \right) \label{beta}\\ &&\alpha^3 \equiv  27 h_1 h_2 h_3 > \frac{1}{2} \beta^2  \label{alpha}   \; .
\eea
To see this, we transform to new coordinates $(v, \psi',r,  \theta, \phi)$ defined by
 \begin{align}
 \td t &= \td v + \left( \frac{A_0}{r^2} + \frac{A_1}{r} \right) \td r\;,  \nonumber \\ 
  \td \psi+  \td \phi &=\td \psi'+ \frac{B_0}{r} \td r    \label{GNC}
 \end{align}
 For a suitable choice of constants $A_0, A_1, B_0$ the spacetime metric and its inverse are analytic at $r=0$. Therefore, the spacetime can be extended to the region $r<0$. The surface $r=0$ is an extremal Killing horizon with respect to the supersymmetric Killing vector $V= \partial /\partial v$.  Near the horizon the scalars $X^i = \alpha/(3h_i)+O(r)$ are regular and the gauge fields are
 \begin{align}
A^i &= \left( \frac{1}{3h_i} + O(r) \right) r\td v   +O(r^2)  \td \phi + \frac{1}{4} k^i \td \psi' \nonumber \\ &  - \left( \frac{\beta}{6h_i} +O(r) \right) (\td \psi'+ 2 \cos \theta \td \phi) \nonumber \\
&+ \left[ \frac{1}{3h_i} \left( A_0 - \frac{\beta B_0}{2} \right) + \frac{1}{4} B_0 k^i + O(r) \right]  \frac{\td r}{r}  \; ,\label{NHgauge}
\end{align}
which shows the only singular terms are pure gauge.  The near-horizon geometry is locally isometric to that of the BMPV black hole~\cite{Breckenridge:1996is, Reall:2002bh}. However, globally the horizon geometry is a lens space $L(2,1) = S^3 / \mathbb{Z}_2$.   To see this, consider the induced metric on cross-sections of the horizon
\begin{equation} \td s^2_3 = \frac{ \alpha^3 - \frac{1}{2} \beta^2}{2\alpha^2}  ( \td\psi' + 2 \cos\theta \td \phi)^2 + 2 \alpha \, \td \Omega_2^2  \; . \label{horizon} 
\end{equation}
Above we showed that asymptotic flatness and smoothness at the centre require $\Delta \psi' = 4\pi$, so (\ref{horizon}) extends to a smooth metric on  $L(2,1)$ as claimed.

It remains to examine regularity and causality in the domain of outer communication (DOC) $r>0$. 
In the Appendix we prove that (\ref{ineq1}), (\ref{h0ineq}) imply that $H H_i>0$ and, remarkably, that this ensures the scalars, the Maxwell fields, the spacetime metric and its inverse are all smooth everywhere in the DOC. Numerical checks also show that the spacetime is stably causal ($g^{tt}<0$) everywhere in the DOC.

\subsection{Geometry of domain of outer communication}
\label{sec:DOC}

Our spacetime has a DOC with non-trivial topology. There is a non-contractible disc $D$ on the axis $\theta=0, 0<r<a$ which degenerates at $r=a$ and ends on the horizon $r=0$,  as we will now show. 

The solution has $U(1)^2$-rotational symmetry. The topology of the spacetime is determined by this $U(1)^2$-action and its fixed points. The $z$-axis of the $\mathbb{R}^3$ base in the Gibbons-Hawking space corresponds to the axes where the $U(1)^2$ Killing fields vanish.  Due to our choice of harmonic functions, the $z$-axis splits naturally into three intervals $I_+ = \{ z >a \} ,\; I_D = \{ 0<z<a \}, \; I_- = \{ z<0 \}$. 
The semi-infinite intervals $I_\pm$ correspond to the two axes of rotation that extend out to infinity. The finite interval $I_D$ corresponds to a non-contractible disc topology surface $D$ that ends on the horizon.  

To see this, consider the geometry induced on the $z$-axis. The 1-forms restrict to $\chi|_{I_\pm}= \pm \td \phi$, $\chi|_{I_D}= 3 \td \phi$ and $\hat{\omega}=0$. Hence, on $I_+$ the Killing field $v_+ = \partial_\phi-\partial_\psi$ vanishes, whereas $\partial_\psi$ is non-vanishing and degenerates smoothly at $z=a$.  Next, on $I_D$ the Killing field $v_D   = \partial_\phi - 3 \partial_\psi$ vanishes, whereas $\partial_\psi$ is non-vanishing even at the horizon end $z \to 0$ and degenerates smoothly at $z=a$.  Thus, the interval $I_D$ corresponds to a surface of disc topology $D$. Lastly, on $I_-$ the Killing field $v_- = \partial_\phi+\partial_\psi$ vanishes and  $\partial_\psi$ is non-vanishing.  Observe that $v_D = 2v_+ - v_-$ and hence in the $2\pi$-normalised $U(1)^2$-basis $( v_+, v_-)$ we may write
\be
v_+ = (1,0), \qquad v_D = (2,-1), \qquad v_- = (0,1)  \; .
\ee
Thus, 
\be
\det ( v_D^T v^T_+ ) =  1    \label{junction}
\ee
and hence the compatibility requirement for adjacent intervals is obeyed~\cite{Hollands:2007aj}. The interval structure is summarized in the figure below.    
\begin{figure}[h]\label{Fig1}
   \centering
    \includegraphics[scale=1.2]{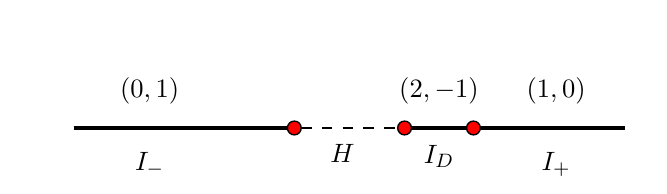} 
  \caption{Interval structure for the black lens metric.}
  \end{figure}
    
The supersymmetric Killing field $V=\partial /\partial t$ may become null in the DOC of the black hole.  Indeed, this is precisely why the black lens evades the uniqueness theorem for the BMPV black hole~\cite{Reall:2002bh}. This `ergosurface' is a timelike hypersurface defined by $f=0$.  Our regularity analysis shows that the zeros of $f$ coincide with those of $H$.  In Cartesian coordinates on the Gibbons-Hawking space, the equation $H=0$ is
\be
x^2+y^2 + (z-2a) \left( z-\frac{2 a}{3} \right)=0  \; ,
\ee
which shows that $\tfrac{2}{3} a \leq  z \leq 2a$ and the endpoints occur only on the axis.  In the spacetime the ergosurface is smooth with topology $\mathbb{R}_t \times S^3$. We may see this as follows.  The metric induced on the axis is regular everywhere including at $z=\tfrac{2}{3} a, 2a$ which correspond to $\mathbb{R}_t \times S^1$ submanifolds.  In particular, the ergosurface is characterised by  $v_D=0$ at $z = \tfrac{2}{3} a$,  $v_+=0$ at $z=2a$ and the $U(1)^2$-acting freely for $\tfrac{2}{3} a <  z < 2a$. Hence, (\ref{junction}) implies the spatial topology of the ergosurface is $S^3$ as claimed.

\subsection{Physical quantities}

The asymptotic electric charges and angular momenta (in units where the 5d Newton constant $G_5=1$) are
\bea
Q_i &=& 3\pi \left(\ell_i+\frac{1}{24} |\epsilon_{ijk}| k^j k^k \right)\\
J_\phi &=& - \frac{3}{2} \pi a \lambda_i k^i  \\
 J_\psi &=& -\pi \left[ \frac{3}{2}(\ell_i+a \lambda_i) k^i + \frac{|\epsilon_{ijk}|}{24}k^ik^jk^k \right]  \; ,
\eea
with the mass given by the BPS condition $M= Q_1+Q_2 +Q_3$. Observe that (\ref{ineq1}) and (\ref{h0ineq}) imply $Q_i>0$ for all $i=1,2,3$.

Inspecting the asymptotic expansions of the gauge field components $A^i_\psi, A^i_\phi$ near infinity reveals that $k^i$ generate a magnetic dipole (the angular momenta also contribute to this).   It is natural to ask if the magnetic dipoles can be expressed as a magnetic flux over some 2-cycle in the spacetime, as in the case of a black ring. The natural candidate is the magnetic flux through the disc $D$,
\be
\Pi_i[D] \equiv \frac{1}{2\pi} \int_D F^i  = -\frac{1}{2} k^i + \frac{h_i \beta}{\alpha^3}  \; .
\ee
Thus this flux does not capture the dipole field alone (note the second term is missing in \cite{Kunduri:2014kja}).

An intrinsic definition of the dipole charge may be obtained as follows.
The Killing field $v_D$ vanishes on $D$ and hence the magnetic potentials $\Phi_i$ defined by $i_{v_D} F^i = \td \Phi_i$ are constant on $D$. For our solution the potentials on $D$, defined to vanish at infinity, are
\be 
q_i \equiv \frac{1}{2} \Phi_i|_D = - \frac{1}{2} k^i  \label{dipole}
\ee
as required (the normalisation is chosen for later convenience). Indeed, these potentials appear in the first law of black hole mechanics as  extensive variables and are the analogues of the dipole charges of a black ring~\cite{Kunduri:2013vka}.

To summarise, we have constructed a five-dimensional solution which is asymptotically flat, regular on and outside a horizon of spatial topology $L(2,1)$.   Thus our solution is a black lens.  The solution is a seven-parameter family specified by $(a, k^i, \ell_i)$, subject to the inequalities (\ref{ineq1}, \ref{h0ineq}, \ref{alpha}).  Equivalently, we may parameterise the solution by the physical quantities $(Q_i, q_i, J_\psi, J_\phi)$ subject to the constraint
\be
J_\psi - J_\phi = q_i \left( Q_i- \frac{\pi}{6} |\epsilon_{ijk}| q_j q_k \right)    \label{constraint}
\ee
and inequalities corresponding to (\ref{ineq1}, \ref{h0ineq}, \ref{alpha}).
 The special case $Q_i=Q$ and $q_i = q$ reduces to the supersymmetric black lens of minimal supergravity (albeit in a simpler parameterisation here)~\cite{Kunduri:2014kja}.
 
 It is worth noting that our regularity constraints (\ref{ineq1}) and (\ref{h0ineq}) imply that $q_i >0$ (or $<0$) for all $i=1,2,3$. Since the dipoles all have the same sign we must have $J_\phi \neq 0, J_\psi \neq 0$ and hence this black hole never has the same asymptotic charges as the BMPV black hole which has $J_\phi=0$ (or $J_\psi=0$). 
 
 We can express the area solely in terms of the physical quantities:
\begin{align}
 A_5 &= 16\pi^2 \left[2 \prod_{i=1}^3 \left(\frac{Q_i}{\pi} - \frac{|\epsilon_{ijk} |q_j q_k}{4}\right)\right. \nonumber \\&\phantom{= 16\pi^2 22}   - \left. \frac{1}{4}\left(\frac{J_\psi + J_\phi}{\pi} - q_1 q_2 q_3\right)^2\right]^{1/2}  \; .
 \end{align} 
In the limit $J_\phi \to 0$ this does not reduce to the area of the BMPV black hole, which in our conventions is 
\begin{align}
A_{\text{BMPV}} = 16 \pi^2 \sqrt{ \frac{Q_1Q_2 Q_3}{\pi^3} - \frac{J_\psi^2}{4\pi^2}}  \; .
\end{align}
 
 \section{D1-D5-P solution}
 \label{sec:D1D5P}
 
 \subsection{Structure and physical properties}
The black lens solutions we have constructed can be uplifted on $T^6$ to yield solutions of eleven-dimensional supergravity.   Via a series of dualities one can map these to D1-D5-P solutions of Type IIB supergravity as in~\cite{Elvang:2004ds}.  In terms of the 5d data, the string frame solution is
 \begin{align}
 \td s_{10}^2 &= (X^3)^{1/2} \td s^2_5 +  (X^3)^{-3/2} (\td z+ A^3)^2 \nonumber  \\  \nonumber& \phantom{=}+ X^1 (X^3)^{1/2} \td z^i \td z^i \; , \qquad e^{2\Phi} = \frac{X^1}{X^2} \\
 F^{(3)} &= (X^1)^{-2} \star_5 F^1+ F^2 \wedge (\td z + A^3)  \; ,   \label{D1D5P}
 \end{align}
 where $z$ is a coordinate on $S^1$ and  $(z^{i}: i=1,2,3,4)$ are coordinates on a flat $T^4$.  We will take the periods of $z$ and $z^i$ to be $2\pi R_z$ and $2\pi L$ respectively. Generically such solutions describe an intersection of D1 and D5 branes carrying momentum P in the $z$ direction, where the  D1 and D5 wrap the $(z)$ and $(z1234)$ directions respectively.

Since we already checked the five-dimensional metric, scalars and Maxwell fields are smooth on and outside an event horizon at $r=0$, the only source of potential singularities in the 10-dimensional geometry comes from the terms involving the gauge field $A^3$. Equation (\ref{gaugecentre}) shows that there exists a gauge in which $A^3$ is smooth at $r_1=0$.  In this gauge the 10-dimensional solution is manifestly smooth at the centre $r_1=0$. 
 
 Inspecting the near-horizon gauge fields (\ref{NHgauge}) reveals that if we define a new coordinate $z'$ by
 \begin{align}
 \td z'  &= \td z + \frac{1}{4}k^3  \td \psi'  \nonumber \\ &+ \left[ \frac{1}{3 h_3} \left( A_0 - \frac{\beta B_0}{2} \right) + \frac{1}{4} B_0 k^3 \right]  \frac{\td r}{r} 
\end{align}
then  $\td z + A^3$ is smooth at $r=0$.  Therefore, the 10-dimensional solution in the coordinates $(v,\psi', r, \phi, \theta, z', z^i)$ is smooth at the surface $r=0$. 
As in five-dimensions the surface  $r=0$ is an extremal Killing horizon with respect to  $V = \partial /\partial v$. However, the gauge which makes $A^3$ regular at the centre $r_1=0$ is not the same as that which makes it regular at $r=0$. In the gauge regular at $r_1=0$  the change of coordinate is
\be
z' = z + \frac{1}{2} k^3\Psi  + O ( \log r)  \; .
\ee
Since $z$ parameterises a circle of radius $2\pi R_z$, requiring the Kaluza-Klein fibration to be globally defined places a quantization condition. We deduce the dipole (\ref{dipole})
\be
q_3 = n_{\text{\tiny KK}}  R_z  \label{KKquant}
\ee
 is quantized  where $n_{\text{\tiny KK}} \in \mathbb{Z}$. This is also consistent with the solution being asymptotically $
 \mathbb{R}^{1,4}\times T^5$ as $r\to \infty$.
 
The near-horizon geometry can be deduced from the five-dimensional one.  Globally it is isometric to $L(2,1)\times T^4$ fibered over the near-horizon geometry of the extremal BTZ black hole. To untwist the fibration define $\psi'' = \psi' - z'\beta/(9 h_1 h_2) $
which gives
\begin{align}
\label{NHG}
&\td s^2_{\text{NH}}=  \frac{ 6 \sqrt{ 2h_1 h_2} \td v \td r}{ \sqrt{\alpha^3 - \tfrac{1}{2} \beta^2}} + \frac{2 r \td v \td z'}{3 \sqrt{h_1 h_2}} + \frac{(\alpha^3 - \tfrac{1}{2} \beta^2)\td z'^2 }{27 (h_1 h_2)^{3/2} } \nonumber \\ 
&+ 6 \sqrt{h_1 h_2} \left[ \frac{1}{4} ( \td \psi''+2 \cos \theta \td \phi)^2 + \td \Omega_2^2 \right]  +  \sqrt{\frac{h_2}{h_1}} \td z^i \td z^i 
\end{align}
The first line is the near-horizon geometry of the extremal BTZ black hole and the second is $L(2,1) \times T^4$.  

In  string theory, the number of D1 branes, D5 branes and units of momentum are
 \begin{align}
&N_{1}  = \frac{4 L^4 Q_2}{\pi g_s \ell_s^6}\;, \; \;  \; N_{5} =   \frac{4 Q_1}{ \pi g_s \ell_s^2}  \; , \; \; \;
N_{\text{\tiny P}} = \frac{4 L^4 R_z^2 Q_3}{\pi g^2_s \ell_s^8}  ,\label{branenumber}
 \end{align}
 and the D1 and D5  quantized dipoles  are
 \begin{eqnarray}
  n_{1} = \frac{L^4 R_z q_1}{g_s \ell_s^6} \; , \qquad
 n_{5} =  \frac{R_z q_2}{g_s \ell_s^2}  \; ,  \label{branedipole}
 \end{eqnarray} 
 where $g_s$ and $\ell_s$ are the string coupling and length. Quantization of $n_1, n_5$ follows from (\ref{KKquant}) and by applying a U-duality transformation which permutes $(n_1, n_5, n_{\text{\tiny KK}})$.

To compute the entropy of our black D1-D5-P system we need the area of the spatial geometry of the horizon in the Einstein frame, $S_{\text{\tiny BH}}= A_{10}/(4G_{10})$, where $16 \pi G_{10} = (2\pi)^7 \ell_s^8 g_s^2$.  We may write this purely in terms of the brane numbers (\ref{branenumber}) and dipoles (\ref{branedipole}):
\begin{align}
S_{\text{\tiny BH}} &= 2\pi \left[2(N_1 - 2n_1 n_{\text{\tiny KK}})(N_5 - 2n_5 n_{\text{\tiny KK}})(N_{\text{\tiny P}} - 2n_1 n_5 )  \right.  \nonumber \\ &\phantom{2\pi 2 \pi 2} \left. - (J_\psi  + J_\phi - 4 n_1 n_5 n_{\text{\tiny KK}})^2 \right]^{1/2}   \; .\label{BHentropy}
\end{align} 
Furthermore, (\ref{constraint}) becomes
\be
J_\psi- J_\phi=  n_1 N_5 + n_5 N_1+ n_{\text{\tiny KK}} N_{\text{\tiny P}} - 4 n_1 n_5 n_{\text{\tiny KK}}   \; ,\label{branecons}
\ee
resulting in a constraint on the quantum numbers.

\subsection{Decoupling limit}
Now consider the decoupling limit of our D1-D5-P solution. This is defined by $\alpha' =\ell_s^2 \to 0$ with $g_s$ and $N_{1}, N_{5}, N_{\text{\tiny P}}, n_{1}, n_{5}, n_{\text{\tiny KK}}$ all  held fixed, such that the energy of the excitations (in string units) near the `core'   $r /\ell_s^4, a/ \ell_s^4$ remain finite. This decouples the bulk geometry from the asymptotically flat region. Further, we keep $R_z$ fixed so that only the momentum modes are the lowest surviving excitations.  On the other hand, we scale the $T^4$  so $z^i/ \ell_s, L / \ell_s$ are fixed so the energies of its excitations are large.    We find that upon an appropriate rescaling of the IIB solution, the decoupling limit is identical to our original solution except $\lambda_1=\lambda_2=0$.   

The decoupling limit inherits all the properties of our original solution and only differs in the asymptotic region $r \to \infty$.  Setting $r= \rho^2/4$, then as $\rho \to \infty$ 
 \begin{align}
 \td s_{\text{\tiny DL}}^2 &\sim \left( \frac{Q_1 Q_2}{\pi^2} \right)^{-\tfrac{1}{2}} \frac{\rho^2}{4} ( - \td t^2+ \td z_\infty^2) \nonumber \\ &+ \left( \frac{Q_1 Q_2}{\pi^2} \right)^{\tfrac{1}{2}}\frac{ 4 \td \rho^2}{\rho^2}+ \sqrt{\frac{Q_2}{Q_1}} \td z^i \td z^i \nonumber \\
 &+ \left( \frac{Q_1 Q_2}{\pi^2} \right)^{\tfrac{1}{2}} \left[ ( \td \psi+ \cos \theta \td \phi)^2 + \td \Omega_2^2 \right] 
 \end{align}
 where $z_\infty= z+t - n_{\text{\tiny KK}} R_z \psi$. This is asymptotically global AdS$_3 \times S^3 \times T^4$ with the radii of AdS$_3$ and $S^3$ both equal to $\tilde{\ell}^2 = 4  \sqrt{Q_1 Q_2/ \pi^2 }$.  By the AdS/CFT duality we thus expect an equivalent description in terms of a 2d CFT with a Brown-Henneaux central charge $c= 3 \ell /2 G_3$ \cite{Brown:1986nw}, where $\ell$ is the AdS$_3$ radius and $G_3$ is the effective 3d Newton constant obtained by a KK reduction on $S^3\times T^4$, all computed in the Einstein frame (using the asymptotics of the dilaton $e^{2\Phi}\sim Q_2/Q_1$).  
 In terms of the brane numbers the central charge is $c= 6 N_1 N_5$, as of course is expected for the D1-D5 CFT.
 
 It is important to note that the decoupling limit is not a product space with a locally AdS$_3$ factor.  It is a non-trivial interpolation between an asymptotically global AdS$_3 \times S^3 \times T^4$ and a near-horizon geometry that is a twisted near-horizon extremal BTZ $\times L(2,1) \times T^4$ given by (\ref{NHG}).  Therefore, in order to apply AdS$_3$/CFT one would have to account for the tower of KK states on $S^3$ that arise from dimensional reduction to 3d~\cite{Skenderis:2008qn}.
 Nevertheless, due to the locally AdS$_3$ factor in the {\it near-horizon geometry} of our D1-D5-P solution, its entropy  can be accounted for by Cardy's formula for the degeneracy of states in the IR CFT~\cite{Strominger:1997eq} (see also \cite{Balasubramanian:2009bg}). In the near-horizon geometry (\ref{NHG}) the AdS$_3$ and $L(2,1)$ radii are both $\tilde{\ell}^2 = 24 \sqrt{h_1 h_2}$.  
Dimensional reduction on $L(2,1) \times T^4$ (in the Einstein frame) leads to 3d Einstein gravity with a Brown-Henneaux central charge
\begin{equation}
c= 12(N_1 - 2 n_1 n_{\text{\tiny KK}})(N_5 - 2n_5 n_{\text{\tiny KK}}) \;  \label{centralc}
\end{equation} 
for the IR CFT.

\subsection{Spectral flow to a black ring}
The asymptotically flat supersymmetric black ring  can also be expressed as a 2-centred Gibbons-Hawking solution with harmonic functions~\cite{Gauntlett:2004qy}
\bea
&&\tilde{H} = \frac{1}{r_1}, \qquad \tilde{K}^i = \frac{\tilde{q}^i}{r},  \nonumber \\  &&\tilde{L}_i = \lambda_i + \frac{\tilde{\ell}_i}{r},\qquad  \tilde{M} = \frac{3 \lambda_i \tilde{q}^i}{4} \left(1- \frac{a}{r} \right) \; ,  \label{ring}
\eea
where we have shifted the horizon to the origin of $\mathbb{R}^3$. Sufficient conditions for regularity of the black ring are the dipoles $\tilde{q}^i > 0$, $\tilde{\ell}_i > 0$ and positivity of the horizon area (which also eliminates CTCs) \cite{Elvang:2004ds}. It can also be uplifted to a D1-D5-P solution (\ref{D1D5P}). Similarly to the black lens, its decoupling limit given by $\lambda_i = (1/3) \delta_{i}^3$, is a non-trivial interpolation between a global AdS$_3 \times S^3 \times T^4$ and a twisted  near-horizon extremal BTZ $\times L(\tilde{n}_{\text{\tiny KK}},1) \times T^4$ where $\tilde{q}^3 = \tilde{n}_{\text{\tiny KK}} R_z$. 

In fact, as we now show, the decoupling limit of the $ n_{\text{\tiny KK}}=1$ black lens is related to a black ring by spectral flow and certain gauge transformations. In 10d these transformations are diffeomorphisms generated by~\cite{Bena:2008wt}
\bea
&&S_\gamma: \psi \mapsto \tilde{\psi} = \psi+ \gamma \frac{z}{R_z} , \\ &&G_g: z\mapsto \tilde{z}= z+ g R_z \psi  \; ,
\eea
where $\gamma \in 2 \mathbb{Z}$ and $g \in \tfrac{1}{2} \mathbb{Z}$ are required for the transformation to be globally defined. These generate an $SL(2,\mathbb{Z})$ symmetry acting on the torus with coordinates $(\psi, z)$. Being diffeomorphisms such transformations must preserve the horizon topology and hence the black ring must have $\tilde{n}_{\text{\tiny KK}}=2$.   

Explicitly, in terms of the harmonic functions 
 \begin{align}
S_\gamma:  \;  &H \to H -\frac{\gamma}{2 R_z}  K^3,\quad K^{i}\to K^{i} + \frac{6\gamma}{R_z} |\epsilon_{ij3}| L_{j}, \nonumber \\ & L_i \to L_i- \frac{2}{3} \frac{\gamma }{R_z} M \delta^3_i , \quad M \to M  \; ,\\
G_g:  \;  &H \to H, \quad K^i\to K^i-2g R_z \delta^i_3 H,  \\ &L_i \to L_i +\frac{1}{6} g R_z |\epsilon_{ij3}| K^{j},  \quad M \to M - \frac{3}{2} g R_z L_3  \; .  \nonumber
\end{align}
It can be shown that the most general $SL(2,\mathbb{Z})$ transformation generated by $S_\gamma$ and  $G_g$, which maps the decoupling limit of the black lens (\ref{blens}) to that of the black ring (\ref{ring}), is $G_{-1} S_2 G_{-1}$, where $ k^3= - 2R_z$ and $\tilde{q}^3 = - k^3$ (we fix $\tilde{q}^3>0$). The KK dipole quantization  condition (\ref{KKquant}) thus gives $n_{\text{\tiny KK}}=1$. Writing this in terms of the black ring KK dipole charge we find $\tilde{n}_{\text{\tiny KK}}= 2 $  as expected.  The rest of the parameters are related by $ \tilde{q}^{1} =- 24 h_1/k^3,  \tilde{q}^{2} =- 24 h_2/k^3$ 
 and  $\tilde{\ell}_i = - \ell_i - 2a \lambda_i$.  
 
 To fully check the map one also has to examine the constraints on the parameters from global regularity and causality.  The inequalities \eqref{ineq1} and \eqref{h0ineq}  are equivalent to $\tilde{\ell}_i + a \lambda_i >0, \;\tilde{q}^{1} > 0, \; \tilde{q}^{2} > 0$ and
\begin{eqnarray}
&&\tilde{\ell}_3 < -2a\lambda_3 + \frac{24}{(\tilde{q}^3)^2} \left( \tilde{\ell}_2+ \tfrac{1}{24}\tilde{q}^1 \tilde{q}^3 \right) \left( \tilde{\ell}_1+ \tfrac{1}{24}\tilde{q}^2\tilde{q}^3 \right)  \; ,  \nonumber
\end{eqnarray} 
whereas (\ref{alpha}) is equivalent to the condition for the absence of CTCs in the black ring spacetime. Thus the regularity and causality constraints for the black lens are consistent with those for the black ring; in fact, apart from the bound on $\tilde{\ell}_3$ they agree precisely. 

The quantized charges of the black ring and black lens are related by
\begin{align}
&\tilde{N}_1= N_1, \qquad \tilde{N}_5 = N_5, \nonumber  \\
& \tilde{N}_{\text{\tiny P}} = - N_{\text{\tiny P}}+ 4 n_1 n_5 +(N_1 - 2 n_1)(N_5 - 2n_5) - 2 J_\phi  \; ,\nonumber \\
&\tilde{n}_1= N_1 - 2 n_1, \qquad \tilde{n}_5 = N_5 - 2n_5 \nonumber \\
&\tilde{J}_2 = -(J_\psi + J_\phi) + N_1 N_5\,, \qquad \tilde{J}_1 = \tilde{J}_2 + 2 J_\phi  \; ,\label{branemap}
\end{align}
where $\tilde{J}_1$ and $\tilde{J}_2$ are the angular momenta along the $S^1$ and $S^2$ of the ring respectively \cite{Elvang:2004ds}.  Using \eqref{branemap} and (\ref{branecons}), it is straightforward to check that the entropy of the $n_{\text{\tiny KK}}=1$ black lens (\ref{BHentropy}) maps to that of the $\tilde{n}_{\text{\tiny KK}}=2 $ black ring (this is of course guaranteed by the map being a diffeomorphism). Also, the IR CFT central charge (\ref{centralc}) maps to that of the black ring $c= 6 \tilde{n}_1 \tilde{n}_5 \tilde{n}_{\text{\tiny KK}}$~\cite{Strominger:1997eq}.

The above shows that we may appeal to the microscopic counting of black ring entropy \cite{Cyrier:2004hj,Bena:2004tk} to supply an account of the entropy for the $n_{\text{\tiny KK}}=1$ subset of black lenses. The microstates of this black lens will be related by the above spectral flow to those of the $\tilde{n}_{\text{\tiny KK}}=2$ black ring. We emphasise though that the above also shows that the $|n_{\text{\tiny KK}}| \neq 1$ black lenses are {\it not} related to a black ring by spectral flow. Thus a microscopic description of the general case remains an open problem.   It would be interesting to derive the entropy of this system directly in terms of the D1-D5 CFT. \\

{\it Acknowledgements.} HKK is supported by NSERC Discovery Grant 418537-2012.  JL is supported by STFC [ST/L000458/1]. We thank Gary Horowitz and Joan Sim\'on for useful comments.  We thank an anonymous referee for the suggestion that the black ring and black lens may be related by spectral flow.

\appendix

\section{Black lenses in $U(1)^N$-supergravity}
\label{appendix}

\subsection{Supersymmetric solutions}
The bosonic sector of five-dimensional $\mathcal{N}=1$ supergravity coupled to $N-1$ abelian vector multiplets consists of a metric $g_{\mu\nu}$, $N$ abelian vectors $A^I$ and $N$ real positive scalars fields $X^I$ subject to the constraint
\begin{equation}
\frac{1}{6} C_{IJK} X^I X^J X^K = 1 \label{cubic}
\end{equation} where $C_{IJK} = C_{(IJK)}$ are real positive constants and the indices $I, J, K, ... =1\ldots N$.  We also define,
\be
X_I = \frac{1}{6} C_{IJK} X^J X^K  \; . 
\ee
The bosonic action is,
\begin{align}\label{generalaction}
S &= \frac{1}{16\pi G_5} \int \left( R \star 1 - G_{IJ} \td X^I \wedge \star \td X^J  \right. \nonumber \\  & \left.- G_{IJ} F^I \wedge \star F^J - \tfrac{1}{6} C_{IJK} F^I \wedge F^J \wedge A^K\right)  \; ,
\end{align} 
where $F^I= \td A^I$ are Maxwell fields and 
\begin{equation}
G_{IJ} \equiv \frac{9}{2} X_I X_J - \frac{1}{2} C_{IJK}X^K  \; .
\end{equation}
We will assume the scalars parameterise a symmetric space so that
\be
\label{symspace}
C_{IJK} C_{J(LM} C_{PQ) K} = \frac{4}{3} \delta_{I (L} C_{MPQ)}   \; .
\ee
This ensures that $G_{IJ}$ is invertible with inverse
\be
G^{IJ} = 2 X^I X^J - 6 C^{IJK} X_K  \; ,
\ee
and 
\begin{equation}\label{Xup}
X^I  =  \frac{9}{2} C^{IJK} X_J X_K  \; 
\end{equation} 
where $C^{IJK} = C_{IJK}$. 

In particular, we will be interested in $U(1)^3$-supergravity which is the special case of this theory when $N=3$ and $C_{IJK} = 1$ if $(IJK)$ is a permutation of $(123)$ and $C_{IJK}=0$ otherwise. Also note that minimal supergravity can be recovered by simply setting $N=1$,  $X^I = \sqrt{3}$ and $C_{111} = 2/ \sqrt{3}$ (note then $X_I= 1/\sqrt{3}$).

A large class of supersymmetric solutions (timelike class) can be written in the canonical form
\begin{equation}\label{bpsmetric}
\td s^2 = -f^2(\td t + \omega)^2 + f^{-1} \td s^2(M_4)
\end{equation} 
where $M_4$ is any a hyperk\"ahler space and $f,\omega$ are a function and 1-form on $M_4$ and $V = \partial /\partial t$ is the supersymmetric Killing field. 
We will take $M_4$ to be a Gibbons-Hawking space
\begin{equation}\label{basemetric}
\td s^2(M_4) = H^{-1} (\td\psi + \chi)^2 + H \td x^i \td x^i
\end{equation} 
where $\chi$ and $H$ are a 1-form and function defined on $\mathbb{R}^3$ obeying $\star_3 \td \chi = \td H$.   
The general local supersymmetric solution with this base is fully determined in terms of  $2N+2$ harmonic functions $H, K^I, L_I, M$ on $\mathbb{R}^3$, as follows \cite{Gauntlett:2004qy}. 

The 1-form  $\omega$ may be decomposed as $\omega = \omega_\psi ( \td\psi + \chi) + \hat{\omega} $ where $\hat{\omega}$ is a 1-form on $\mathbb{R}^3$. It is given by
\begin{equation}
\omega_\psi = -\frac{1}{48} H^{-2} C_{IPQ} K^I K^P K^Q - \frac{3}{4} H^{-1} L_I K^I + M
\end{equation}  and
\begin{equation}
\star_3 d\hat\omega = H d M - M d H + \frac{3}{4}(L_I d K^I - K^I d L_I)
\end{equation}  The scalars are given by,
\begin{equation}\label{X_I}
H_I \equiv f^{-1} X_I =\frac{1}{24} H^{-1} C_{IPQ} K^P K^Q + L_I
\end{equation} which using the constraint \eqref{cubic}
implies that 
\begin{equation}\label{finvcubedH}
f^{-3} = \frac{9}{2}C^{IJK} H_I H_J H_K
\end{equation}
so the function $f$ is also determined.  Finally, the gauge fields are
\be\label{A^I}
A^I = X^I f(\td t + \omega) + \frac{1}{2}  \left( H^{-1} K^I (\td \psi+ \chi) +\xi^I  \right)  
\ee
where $\star_3 \td \xi^I= -\td K^I$.

 Now consider the 2-centred solution given by
 \begin{align}
 H &= \frac{2}{r} - \frac{1}{r_1} , \qquad K^I = \frac{k^I}{r} \label{Harm}
  \; ,  \\   L_I &= \lambda_I + \frac{\ell_I}{r} + \frac{\ell_{1I}}{r_1} ,\qquad M = m +\frac{m_0}{r}+ \frac{m_1}{r_1}    \nonumber
 \end{align}
where $r_1 = \sqrt{r^2+a^2- 2r a \cos \theta }$ is the Euclidean distance from a `centre' in $\mathbb{R}^3$ with Cartesian coordinates $(0,0,a)$  and we assume $a>0$. Integrating gives
 \begin{align}
 \chi &= \left[ 2\cos\theta - \frac{r\cos\theta - a}{r_1} \right] \td\phi  , \quad \xi^I = -k^I \cos \theta  \td \phi ,  \nonumber \\
\hat{\omega} &=\left\{  -\left(2m - \tfrac{3}{4}\lambda_I k^I\right)\cos\theta + \frac{m (r \cos\theta -a)}{r_1}  \right. \nonumber \\
 &\phantom{+}+ \left. \frac{(r-a\cos\theta)\left(m_0+2m_1 - \tfrac{3}{4}\ell_{1I}k^I \right)}{a r_1}  +c  \right\}   \td \phi   \label{1forms}
 \end{align}
where the freedom in $\hat{\omega}, \chi$ and $\xi^I$ has been fixed by shifts in $t, \psi$ and gauge transformations in $A^I$ respectively. 

The spacetime is asymptotically flat $\mathbb{R}^{1,4}$ provided we make the identifications $\psi \sim \psi + 4\pi$,  $\phi \sim \phi + 2\pi$ and $\theta \in [0,\pi]$ and we choose the constants such that
\bea
&&\frac{9}{2} C^{IJM} \lambda_I \lambda_J \lambda_M =1,   \label{lambdaconstr} \\ 
 && m = \frac{3}{4} \lambda_I k^I  \; , \quad c = \frac{3\ell_{1I} k_{0}^I -4m_0 - 8m_1}{4a} \; .
\eea
Indeed, setting $r=\tfrac{1}{4} \rho^2$ these choices ensure that as $\rho \to \infty$
\begin{align}
f = 1 +O(\rho^{-2}), \quad \omega_\psi = O(\rho^{-2}), \quad \omega_\phi= O(\rho^{-2})
\end{align}
and hence asymptotically the spacetime is given by (\ref{AF}).  Further from \eqref{X_I} it is easy to verify that asymptotically $X_I =  \lambda_I+O(\rho^{-2})$
and so we deduce
\be
\lambda_I>0  \; .
\ee
The gauge fields are asymptotically pure gauge
\bea
A^I \sim \lambda^I \td t + \frac{1}{2} k^I \td \psi   \; ,
\eea
 where $\lambda^I = \frac{9}{2} C^{IJK} \lambda_J \lambda_K$ and subleading terms $O(\rho^{-2})$.

\subsection{Regularity analysis}
We now perform a careful regularity analysis of the solutions constructed above. Although the solution appears singular at the `centres' $r_1=0$ and $r=0$ we will show that by a suitable choice of constants $r_1=0$ corresponds to a smooth timelike point in the spacetime  whereas $r=0$ corresponds to a regular event horizon. Furthermore, we will confirm that the solution is regular everywhere else in the DOC $r>0$ including the ergosurface where $f$ vanishes.

 \subsubsection{Smooth centre}
 Here we consider smoothness near the centre $r_1=0$.  It is convenient to introduce spherical polar coordinates $(r_1, \theta_1)$ on $\mathbb{R}^3$ adapted to the centre $(0,0,a)$, where $r_1$ is as above, $r_1 \cos \theta_1= z-a$ and $\phi_1=-\phi$.  Let $\rho_1 = 2 \sqrt{r_1}$ and $\psi_1= \psi -2 \phi_1$.  One finds that
 \begin{align}
 \td s^2(M_4) &= F_1 \left( \td\rho_1^2+ \tfrac{1}{4} \rho_1^2 \left[ \td\theta_1^2 + \sin^2 \theta_1 \td \phi_1^2 \right. \right. \\ &+\left. \left. F_1^{-2}(\td \psi_1 + \cos \theta_1 \td \phi_1 + G_1 \td \phi_1 )^2 \right] \right) \nonumber
 \end{align}
 where we have defined
 \be
 F_1 = \tfrac{1}{4} \rho_1^2 H , \qquad G_1= 2-\chi_\phi - \cos \theta   \; .
 \ee
 It is readily verified that our solution obeys $F_1= -1 + O(\rho_1^2)$ and $G_1= O(\rho_1^4)$ as $\rho_1 \to 0$  so that
 \bea
 \td s^2(M_4) \sim & -& \left( \td\rho_1^2 +  \tfrac{1}{4} \rho_1^2 \left[ \td\theta_1^2 + \sin^2 \theta_1 \td \phi_1^2 \right. \right. \nonumber \\  &+&  \left. \left.  (\td \psi_1 + \cos \theta_1 \td \phi_1 )^2 \right] \right)
 \eea
 which shows that the Gibbons-Hawking space approaches the origin of $-\mathbb{R}^4$, provided we choose the periods of the angles as required by asymptotic flatness.
 
 To investigate smoothness at the centre it is convenient to use plane polar coordinates $(X,\Phi)$ and $(Y,\Psi)$ on orthogonal 2-planes of  $\mathbb{R}^4$. These are given by
 \bea\label{centrechart}
 X &=& \rho_1 \cos ( \tfrac{1}{2} \theta_1) \; , \qquad \Phi =\tfrac{1}{2} ( \psi_1+\phi_1) \;  , \\ Y &=& \rho_1 \sin (\tfrac{1}{2} \theta_2) \; , \qquad \Psi =\tfrac{1}{2} ( \psi_1-\phi_1) \; ,  \label{Psicoord}
 \eea
 so that
 \be
 \td s^2 ( \mathbb{R}^4) = \td X^2 + X^2 \td \Phi^2 + \td Y^2 + Y^2 \td \Psi^2   \; .
 \ee
 Any $U(1)^2$-invariant smooth function on $\mathbb{R}^4$ must be a smooth function of $X^2, Y^2$. We find
 \bea
 F_1 
 = - 1+ \dots  \; ,\quad
 G_1 
 =  8X^2 Y^2 (1+ \dots)  \; ,
 \eea
 where $\dots$ are analytic functions of $X^2,Y^2$ which vanish at $X=Y=0$. Using this we deduce
 \bea
 \td s^2 (M_4) &=& \td s^2(\mathbb{R}^4) + O(X^4) \td \Phi^2 + O(1)  X^2 Y^2 \td \Phi \td \Psi \nonumber  \\&+& O(Y^4) \td \Psi^2  \; ,
 \eea
 with higher-order terms all analytic in $X^2,Y^2$.  This shows the Gibbons-Hawking base metric is smooth (in fact analytic) at the centre $r_1=0$.
 
 Next, we demand that the centre $r_1=0$ to be timelike. Since the invariant $V^2 = -f^2$ this requires that $f$ is smooth and non-vanishing at $r_1=0$. In fact, in order to get the spacetime metric signature correct we need  $f|_{r_1=0} <0$.
 We will also demand that the scalars $X_I$ are smooth positive functions. Thus the functions $H_I = f^{-1} X_I$ must be smooth and negative at the centre.  Using the explicit form of our 2-centred solutions we find that these conditions require
\begin{align}
&\ell_{1I} =0 
  \label{ell1}  , \\
  & \ell_I+ a\lambda_I<0   \; .  \label{ineqaxis1}
\end{align} 
With these conditions $f$ and $X_I$ are in fact analytic functions in $X^2,Y^2$ at the centre.

Next, consider the invariant
\be
| \partial_\psi |^2 = \frac{1}{fH} - f^2 \omega_\psi^2  \; .
\ee
The absence of CTCs requires $| \partial_\psi |^2 \geq 0$. But at the centre $| \partial_\psi |^2|_{r_1=0} = - f^2 \omega_\psi^2$ and therefore  we deduce that $|\partial_\psi|_{r_1=0}=0$ and $\omega_\psi |_{r_1=0}=0$.  Therefore, the Killing field $\partial_\psi$ has a fixed point in the spacetime. Furthermore, the invariant $V \cdot \partial_\psi = - f^2 \omega_\psi$ shows that $\omega_\psi$ must be a smooth function on spacetime at and near the centre. Thus, putting things together we deduce that $\omega_\psi$ is a smooth spacetime function which vanishes at the centre $r_1=0$.  The general form of our 2-centred solution has a $1/r_1$ singular term as $r_1 \to 0$. The condition for its absence is 
\be
m_1   =0
\label{m1}
\ee
where we have used (\ref{ell1}).  Furthermore, the condition $\omega_\psi|_{r_1=0}=0$ reduces to
\begin{eqnarray}
m_0 = -\frac{3}{4} a \lambda_I k^I  \; .
 \label{constr}
\end{eqnarray} 
It can now be verified that these conditions imply that
\be
\omega = O(X^2) \td \Phi + O(Y^2) \td \Psi
\ee
with higher order terms analytic in $X^2,Y^2$. Hence the 1-form $\omega$ is analytic at the centre.  Putting the above together, we have shown that the above conditions on the constants ensure the spacetime metric is smooth (in fact analytic) at the centre.    

We now turn to the gauge fields \eqref{A^I}. The above analysis already shows that $X^I f (\td t + \omega)$ is smooth at the centre. Using the above conditions on the constants,  one can verify that near the centre,
\be
A^I= X^I f \td t+ \frac{1}{2}[ k^I+  O(X^2)] \td \Phi - \frac{1}{2} [ k^I+ O(Y^2) ]\td \Psi 
\ee
with higher-order terms analytic in $X^2,Y^2$.  This shows that the Maxwell fields are smooth at the centre. Furthermore,  there is a gauge choice in which the gauge field is a smooth at the centre.

To summarise, we have shown that the spacetime metric, Maxwell fields and scalars are all smooth at the centre $r_1=0$ if the constants are chosen as above.

 \subsubsection{Event horizon}
 
 Now we consider the centre $r=0$. 
 We will show that in fact it corresponds to a regular event horizon provided  
 \be
 h_I >0, \qquad \alpha^3 - \frac{1}{2} \beta^2  >0 ,   \label{horregular}
 \ee
 where the constants $h_I, \alpha, \beta$ are defined by
  \bea
&&h_I \equiv \ell_I + \tfrac{1}{48} C_{IJK} k^J k^K,   \label{h_I} \\  && \alpha \equiv   \left( \tfrac{9}{2} C^{IJK} h_I h_J h_K \right)^{\tfrac{1}{3}} \; ,\\
&&\beta \equiv \frac{3}{4} k^I \left(  \ell_I+2a \lambda_I + \frac{1}{72} C_{IJK} k^J k^K \right)  \; . \label{beta}
 \eea 
 Observe that $h_I>0$ implies $\alpha > 0$.  
 
 To this end, we transform to new coordinates $(v,r, \psi', \theta, \phi)$ given by \eqref{GNC}
for some constants $A_0, A_1, B_0$ to be determined. This gives
 \begin{align}
 g_{vv} &= - f^2 = - \frac{r^2}{\alpha^2} + O(r^3) \nonumber \\  g_{v \psi'} &= - f^2 \omega_\psi = \frac{\beta}{2 \alpha^2} r + O(r^2) \nonumber \\  
 g_{\psi\psi'} &= - f^2 \omega_\psi^2 + (Hf)^{-1}  = \frac{ \alpha^3 - \frac{1}{2} \beta^2}{2\alpha^2} + O(r)  \; .
 \end{align}
 In general, $g_{rr}$ contains $1/r^2$ and $1/r$ singular terms, whereas $g_{r \psi'}$ contains $1/r$ singular terms. Requiring that the $1/r$ singularity in $g_{r\psi'}$ and the $1/r^2$ singularity in $g_{rr}$ are absent fixes the constants,
 \be
 B_0 = -\frac{\beta A_0}{ \alpha^3 - \tfrac{1}{2} \beta^2}, \qquad A_0^2 = 2 \left(\alpha^3 - \frac{1}{2} \beta^2 \right)  \; .
 \ee
 Furthermore, demanding that the $1/r$ singularity is $g_{rr}$ is absent fixes $A_1$ to be a complicated constant (we do not display it as we will not need it). We now have,
 \begin{align}
g_{r\psi'}  &= O(1), \qquad  g_{rr} = O(1), \\
 g_{vr}  &=- \frac{A_0 \alpha}{\alpha^3 - \tfrac{1}{2} \beta^2}   +O(r)   =  \pm \frac{ \sqrt{2} \alpha}{ \sqrt{ \alpha^3 - \tfrac{1}{2} \beta^2}}  +O(r) \; , \nonumber
\end{align}
where $A_0>0$ corresponds to the lower sign and $A_0<0$ to the upper sign. Finally, to assemble the full metric we will also need $\hat{\omega} = O(r)$ and
 \be
 \chi = (1+ 2 \cos \theta + O(r^2)) \td\phi, \quad g_{\Omega \Omega} = 2 \alpha +O(r)  \; .
 \ee

The metric and its inverse are now analytic at $r=0$ and therefore the spacetime can be extended to the region $r<0$. The supersymmetric Killing field $V = \partial / \partial v$ is null on the hypersurface surface $r=0$ and
\be
V_\mu \td x^\mu |_{r=0}  =  \pm \frac{ \sqrt{2} \alpha}{ \sqrt{ \alpha^3 - \tfrac{1}{2} \beta^2}}   (\td r)|_{r=0}
\ee
which shows that $r=0$ is a Killing horizon of $V$. It is easily seen to be a degenerate horizon. The upper sign corresponds to the future horizon and the lower sign to the past horizon.   

The matter fields are also analytic at the horizon. The scalars  are
\be
X^I = \frac{h^I}{\alpha^2}+O(r)  \; ,
\ee 
where we have defined $h^I=\tfrac{9}{2}C^{IJK}h_J h_K$. 
The gauge fields in the new coordinates are  (for any value of $A_0, A_1, B_0$) 
\begin{align}
A^I &=  \left( \frac{h^I}{\alpha^3} + O(r) \right) r\td v  + \frac{1}{4} k^I \td \psi'  + O(r^2) \td \phi   \nonumber \\
&+ \left[ \frac{h^I}{\alpha^3} \left( A_0 - \frac{\beta B_0}{2} \right) + \frac{1}{4} B_0 k^I + O(r) \right]  \frac{\td r}{r}   \nonumber  \\ &- \left( \frac{h^I \beta}{2\alpha^3} +O(r) \right) (\td \psi'+ 2 \cos \theta \td \phi) 
\end{align}
which shows the only singular terms are pure gauge. Hence the Maxwell fields are analytic at the horizon.

The near-horizon geometry may be extracted by taking the scaling limit $(v,r) \to (v/\epsilon, \epsilon r)$ and $\epsilon \to 0$. The result is
\begin{align}
\td s^2_{\text{NH}} &= - \frac{r^2}{\alpha^2} \td v^2 \pm  \frac{ 2\sqrt{2} \alpha}{ \sqrt{ \alpha^3 - \tfrac{1}{2} \beta^2}}  \td v \td r    \nonumber \\ &+ \frac{\beta}{\alpha^2} r \td v ( \td\psi' + 2 \cos\theta \td \phi) + \td s_3^2  \; , \label{NHmetric}  \nonumber\\
F^I_{\text{NH}} &= \frac{h^I}{\alpha^3}\td \left[ r \td v - \frac{\beta}{2}\left(\td \psi' + 2 \cos\theta d\phi\right)\right]  \; , \nonumber \\
X^I_{\text{NH}} &= \frac{h^I}{\alpha^2} \; .
\end{align}
where $\td s^2_3$ is the metric on spatial cross-sections of the horizon \eqref{horizon}.
This is {\it locally} isometric to the BMPV near-horizon geometry. However, the period of $\psi'$ has been fixed to be $4\pi$ by asymptotic flatness and regularity at the smooth centre. Therefore, cross-sections of the horizon $r=0, v= \text{const}$ are lens spaces $L(2,1)$.  

 \subsubsection{Domain of outer communication}
 
 Now we will examine regularity of the solution in the domain of outer communication (DOC) $r>0$.  It is convenient to define
 \be
 \tilde{H}_I \equiv H H_I = H L_I + \frac{1}{24} C_{IJK} K^JK^K  \; .
 \ee
 Explicitly, we can write
 \begin{align}
 \tilde{H}_I &= \frac{P_I}{r^2r_1}, \qquad \text{where} \\
 P_I &=  2 r_1 h_I +  \lambda_I r  [ 2r_1 - (r-a)]  - r \left( \ell_I+ a\lambda_I   \right) . \nonumber
 \end{align}
The inequalities (\ref{horregular}) and (\ref{ineqaxis1}) and the geometric condition $2r_1 \geq r-a$ thus imply 
 \be
 P_I>0   \label{HIpos}
 \ee
 everywhere in the DOC (including $r_1=0$).
 
We may write the invariant
 \be
 f= \frac{H}{\left[ \frac{9}{2} C^{IJK}\tilde{H}_I \tilde{H}_J \tilde{H}_K \right]^{1/3}}  \; .
 \ee
Using (\ref{HIpos}) we deduce that $f$ is smooth everywhere in the DOC, and therefore the zeroes of $f$ coincide with those of $H$. We can write the scalars as
\be
X_I = \frac{\tilde{H}_I}{\left[ \frac{9}{2} C^{IJK}\tilde{H}_I \tilde{H}_J \tilde{H}_K \right]^{1/3}} 
\ee
which shows that $X_I$ is a smooth positive function everywhere in the DOC.

The metric and inverse metric  can be written as
\begin{widetext}
\begin{align}
g_{tt} &= -f^2  \; ,\qquad g_{t\psi} = - f^2 \omega_\psi =\frac{ \frac{1}{48} C_{IPQ} K^I K^P K^Q + \frac{3}{4} H L_I K^I -  H^2M}{\left[ \frac{9}{2} C^{IJK}\tilde{H}_I \tilde{H}_J \tilde{H}_K \right]^{2/3}} , \qquad g_{ti} = -f^2\hat{\omega}_i +g_{t\psi} \chi_i  \nonumber \\
g_{\psi \psi} &= f^{-1} H^{-1} - f^2 \omega_\psi^2 \nonumber \\ \nonumber
&=   \frac{\frac{9}{16}C^{IJM}C_{IPQ}K^P K^Q L_J L_M + \frac{9}{2} H C^{IJK}L_IL_J L_K   - M^2 H^2 - \frac{9}{16} (L_I K^I)^2 - \frac{1}{24} M C_{IJK}K^I K^J K^K - \frac{3}{2} H M L_I K^I}{\left[ \frac{9}{2} C^{IJK}\tilde{H}_I \tilde{H}_J \tilde{H}_K \right]^{2/3}}  \\
g_{\psi i} &=g_{\psi\psi} \chi_i +g_{t\psi} \hat{\omega}_i  \; , \qquad 
g_{ij} = f^{-1} H \delta_{ij} + g_{\psi\psi} \chi_i \chi_j - f^2 \hat{\omega}_i \hat{\omega}_j + 2 g_{t\psi}\chi_{(i} \hat{\omega}_{j)}  \nonumber  \; , \\
g^{tt} &=- Hf^{-1}g_{\psi\psi} +f H^{-1}\hat \omega^2 \; , \qquad 
g^{t\psi} = Hf^{-1} g_{t\psi} + f H^{-1} \hat{\omega}_i \chi_i , \; , \qquad g^{t i} = - f H^{-1} \hat{\omega}^i \; , \nonumber  \\ g^{\psi i} &= - f H^{-1} \chi^i \; , \quad
g^{\psi\psi} = fH + f H^{-1} \chi_i \chi_i \; ,\qquad g^{ij} = f H^{-1} \delta^{ij}   \; , \quad \det g_{\mu \nu} = - H^2 f^{-2}
\end{align} 
\end{widetext}
where we used \eqref{finvcubedH} and \eqref{symspace} to simplify $g_{\psi\psi}$. 
By inspection it is clear that $\chi_i$ and $\hat{\omega}_i$ are smooth in the DOC everywhere except at $r_1=0$. Therefore, remarkably,  (\ref{HIpos}) also ensures that all metric and inverse metric components are smooth everywhere except possibly $r_1 = 0$.  Above we showed the spacetime is in fact smooth at $r_1=0$ and hence we deduce that the metric and inverse metric are smooth everywhere in the DOC.
\begin{widetext}
Finally, the gauge field components are
\begin{align}
A^I_t & = \frac{C^{IJK} \tilde{H}_J \tilde{H}_K  H}{C^{PQR}\tilde{H}_P \tilde{H}_Q \tilde{H}_R}  \;, \qquad A^I_i = A^I_\psi \chi_i + \frac{1}{2} \xi^I_i  \\
 A^I_\psi &= \frac{1}{\left[ \frac{9}{2} C^{LMN}\tilde{H}_L \tilde{H}_M \tilde{H}_N \right]} \left[ \frac{9}{2} C^{IJK} \tilde{H}_J \tilde{H}_K \left(HM - \frac{3}{4} L_IK^I \right) +\frac{9}{32} H K^I C^{KJM} C_{KPQ} L_J L_M K^P K^Q \right.  \nonumber \\
 & \left. +\frac{1}{128}\left(2K^I L_J K^J - C^{IJK} C_{KMN} L_J K^M K^N - 12 H C^{IJK} L_J L_K\right)C_{PQR}K^P K^Q K^R + \frac{9}{4} H^2 K^I C^{PQR}L_P L_Q L_R \right] \nonumber
\end{align}
\end{widetext}
Therefore (\ref{HIpos}) also guarantees the gauge field is smooth everywhere in the DOC except at $r_1=0$. Above we showed that at $r_1=0$ the only singular terms are pure gauge and hence we deduce the Maxwell fields are smooth in the DOC.

We also require our spacetime to be stably causal in the DOC $g^{tt}<0$.  We have verified this numerically in the case of $U(1)^3$-supergravity and find that no further conditions on the parameters need to be imposed.

The geometry and topology of the DOC is discussed in section \ref{sec:DOC}.

\subsection{Physical quantities}

We have constructed an asymptotically flat solution which is regular everywhere on and outside an event horizon of spatial topology $L(2,1)$. 
Our solution is parameterised by the constants $(\lambda_I, \ell_{I} , k^I, a)$ subject to the constraint (\ref{lambdaconstr}), resulting in a $3N$ parameter family of solutions. Furthermore, these parameters obey the inequalities $a>0$, (\ref{ineqaxis1}) and (\ref{horregular}).

The electric charges associated to the Maxwell fields $F^I$ are defined by
\begin{equation}
Q_I = \frac{1}{8\pi} \int_{S^3_{\infty}} G_{IJ} \star F^J  \; . 
\end{equation} 
We find
\begin{align}
Q_I = 3 \pi \left(  \ell_I  + \frac{1}{24} C_{IJK} k^J k^K \right)
\end{align}
where we have used the symmetric space condition (\ref{symspace}) to simplify the expression.  The mass saturates the BPS bound $M = \lambda^I Q_I$. The angular momenta  are
\begin{align}
J_\phi &= -\frac{3 \pi a \lambda_I k^I }{2}\\
J_\psi &= -\pi \left[\frac{3}{2}(\ell_I+ a\lambda_I) k^I + \frac{C_{IJK}}{24}k^Ik^Jk^K \right]   \; .
\end{align} 
It should be noted that $J_\psi, J_\phi$ are the angular momenta with respect to the Euler angles of the $S^3$ at infinity. The angular momenta with respect to the orthogonal $U(1)^2$ angles at infinity, $\phi_1= (\psi+\phi)/2$ and $\phi_2= (\psi- \phi)/2$, are obtained by $J_1= J_\psi+ J_\phi$ and $J_2= J_\psi- J_\phi$.

The asymptotic expansions of the gauge fields in terms of the orthogonal $U(1)^2$ angles at infinity are
\begin{align}
A^I_{\phi_1} &\sim\frac{1}{2} k^I + \frac{4 \cos^2 \left(\frac{\theta}{2} \right)}{\rho^2}\left( \frac{\lambda^I J_{1}}{\pi} +  k^I a \right) \\
A^I_{\phi_2} &\sim\frac{1}{2} k^I + \frac{4 \sin^2 \left(\frac{\theta}{2} \right)}{\rho^2}\left( \frac{\lambda^I J_{2}}{\pi} -  k^I a \right)
\end{align}
Thus the $k^I$ generate a magnetic dipole field at infinity. As discussed above \eqref{dipole}, the dipole charges may be defined by
\begin{equation}
q^I = \frac{1}{2}\Phi^I[D] = -\frac{k^I}{2}
\end{equation} 
where $D$ is the disc topology surface in the DOC discussed in section \ref{sec:DOC}, and the potentials $\Phi^I[D]$ are defined by $\td \Phi^I = i_{v_D} F^I$ where $v_D = \partial_\phi - 3 \partial_\psi$ vanishes on $D$ and the requirement that $\Phi^I \to 0$  at infinity.  The magnetic flux through $D$ is
\be
\Pi^I[D] = \frac{1}{2\pi} \int_D F^I  = -\frac{1}{2} k^I + \frac{h^I \beta}{\alpha^3} \; .
\ee
The conserved charges and dipole charges satisfy the constraint
\be
J_\psi- J_\phi= q^I Q_I - \frac{\pi}{6} C_{IJK}q^I q^J q^K \;.  \label{physcons}
\ee The area of cross-sections of the horizon is
\begin{widetext}
\begin{align}
A_5 = 16 \pi^2 \left[ \frac{1}{3} C^{IJK} \left( \frac{Q_I}{\pi} - \frac{1}{4}C_{IPP'} q^P q^{P'} \right)  \left( \frac{Q_J}{\pi} - \frac{1}{4}C_{JQQ'} q^Q q^{Q'} \right) \left( \frac{Q_K}{\pi} - \frac{1}{4}C_{KRR'} q^R q^{R'} \right) \right. \\ \left. - \left( \frac{1}{2\pi} (J_\phi+ J_\psi) - \frac{1}{12} C_{IJK} q^I q^J q^K  \right)^2 \right]^{\tfrac{1}{2}}  \;  . \nonumber
\end{align}
\end{widetext}

\end{document}